\documentclass[a4paper,11pt]{article}
\usepackage{pos}
\usepackage{floatrow}

\title{Magnetic anomaly coefficients for QCD axion couplings} 

\author*[a]{Anton V. Sokolov}
\author[a]{Andreas Ringwald}

\affiliation[a]{Deutsches Elektronen-Synchrotron DESY, Notkestr. 85, 22607 Hamburg, Germany}

\emailAdd{anton.sokolov@desy.de}
\emailAdd{andreas.ringwald@desy.de}

\abstract{We argue that there are both experimental and theoretical reasons to reconsider the construction of KSVZ-like axion models. From the experimental side, predictions for the axion-photon coupling are not consistent with the collection of astrophysical hints. From the theoretical side, we find that the construction can be conceptually simplified. In particular, it contains an unnecessary assumption on the type of the gauge theory involved which has far-reaching consequences for low energy axion phenomenology. In order to relax this assumption, we allow heavy fermions of the KSVZ-like models to carry magnetic charge. We then show that the axion-photon coupling is generically dominated by the axial anomaly of magnetic currents. Finally, we derive the expression for magnetic anomaly coefficients, which determine the range of numerical values for the axion-photon coupling.}
 
\FullConference{%
  *** The European Physical Society Conference on High Energy Physics (EPS-HEP2021), ***\\
  *** 26-30 July 2021 ***\\
  *** Online conference, jointly organized by Universität Hamburg and the research center DESY ***
}

\begin{document}
\begin{flushright}
		DESY 21-140\\
\end{flushright}
\maketitle

\section{Motivation}

Axions are pseudo Nambu-Goldstone bosons of the spontaneously broken anomalous global $U(1)_{\text{PQ}}$ symmetry~\cite{Peccei:1977hh, Peccei:1977ur, Weinberg:1977ma, Wilczek:1977pj}. They are one of the best motivated hypothesised particles beyond the Standard model. Not only do axions provide an attractive dynamical solution to the strong CP problem, but they are also perfect candidates to account for the cold dark matter abundance suggested by astrophysical and cosmological observations~\cite{Preskill:1982cy,Abbott:1982af,Dine:1982ah}. There is a vast number of ideas how one can potentially discover axions, both directly in the laboratory and indirectly through astrophysical data. Many of them have been already put into practice and more are yet to be implemented in the future. Although the axion has not been discovered so far, its parameter space has been constrained and even some hints pointing to a particular range of parameters have been found. Since most of the axion searches exploit especially its coupling to photons, it is very important to understand which values of this coupling are theoretically preferred. In the conventionally used axion models, which are KSVZ- and DFSZ-like models~\cite{Kim:1979if, Shifman:1979if, Dine:1981rt, Zhitnitsky:1980tq}, it is predicted that on a plot of axion-photon coupling versus axion mass the relevant values occupy a band, see fig.~\ref{fig1}. There are however reasons, both experimental and theoretical, to question the relevance of this band. From the experimental side, there are astrophysical hints, associated to the cooling of horizontal branch stars in globular clusters~\cite{Ayala:2014pea} and anomalous TeV-transparency of the Universe~\cite{DeAngelis:2008sk, Horns:2012fx, Troitsky:2020rpc}, which agree among each other and after being combined point to a parameter region far above the conventional axion band. From the theoretical side, we find that there is an unjustified assumption entering the construction of the KSVZ-like models, which influences drastically predicted values for the axion-photon coupling. After relaxing this assumption, one arrives at a conceptually simpler model, which we analyse in this work. What we find is that, quite intriguingly, the phenomenology of the latter minimal model does agree with the astrophysical hints mentioned earlier, thus resolving the existing controversy between theory and experiment. 

Our narration proceeds as follows. In sec.~2, we show how one can conceptually simplify the construction of the KSVZ-like models by relaxing an assumption on the type of the gauge theory involved. In sec.~3, we explore the phenomenology of the resulting model by relating axion effective field theory operators to the expression for the axial anomaly in a generic Abelian gauge theory, where both electric and magnetic gauge charges are allowed. 

\section{KSVZ-like models are intrinsically biased}

Normally, construction of the KSVZ-like axion models proceeds as follows. One introduces a complex scalar field $\Phi$, which breaks the $U(1)_{\text{PQ}}$ symmetry spontaneously after relaxing to its non-zero vacuum expectation value $v_a / \sqrt{2}$. For consistency with observations, $v_a$ must correspond to some high energy scale. Moreover, in order to solve the strong CP problem, the $U(1)_{\text{PQ}}$ symmetry has to be color anomalous. This is achieved via introducing a new vector-like colored fermion~$\psi = \psi_L + \psi_R$, so that $U(1)_{\text{PQ}}$ acts differently on the two chirality components of~$\psi$. These requirements lead naturally to the following Lagrangian:
\begin{equation}\label{1}
\mathcal{L} \; \supset \; i\bar{\psi} \gamma^{\mu} D_{\mu} \psi + y \left( \Phi \, \bar{\psi}_L {\psi}_R + \text{h.c.}\, \right) - \lambda_{\Phi} \left( \left| \Phi \right|^{2} -\frac{v_a^2}{2} \right)^{\!\! 2},
\end{equation}
where $y$ and $\lambda_{\Phi}$ are some dimensionless constants and $D_{\mu}$ is a covariant derivative encoding the interaction of $\psi$ with the gauge fields of the Standard model. Note that as a result of the spontaneous symmetry breaking $\psi$ gets a mass $m = yv_a / \sqrt{2}$, which is very large for reasonable values of $y$. What remains as a low energy degree of freedom is a pseudo Nambu-Goldstone boson of the spontaneous symmetry breaking $a$ -- the axion -- which can be thought of as an angular mode of $\Phi$. The axion interacts with the gauge fields of the Standard model through loops of $\psi$. In particular, at low energies, where the relevant gauge bosons are photons and gluons, these interactions are given by the axion field times the expressions for axial electromagnetic and color anomalies, respectively, with some coefficients:
\begin{equation}
\mathcal{L}_a\; \supset \; -\frac{1}{4}\, g^0_{a\gamma \gamma} \,a\, F_{\mu \nu} \tilde{F}^{\mu \nu}\, +\, \frac{a g_s^2}{32\pi^2 f_a}\; G_{\mu \nu}^a \tilde{G}^{a\, \mu \nu} \, , \quad f_a = \frac{v_a}{2N}\, , \quad g^0_{a\gamma \gamma} =  \frac{E}{N}\cdot \frac{e^2}{8\pi^2 f_a}  \, .  
\end{equation}
The parameters $E$ and $N$ are called electromagnetic and color anomaly coefficients, respectively. They depend on the representation of $\psi$ under the gauge symmetries of the Standard model. Since the latter representation is unknown, $E$ and $N$ can in principle vary considerably, which gives rise to an uncertainty in the parameter $g^0_{a\gamma \gamma}$~\cite{DiLuzio:2017pfr}. This uncertainty translates into a band on the plot of possible axion-photon couplings $g_{a\gamma \gamma}$ versus axion mass $m_a$, see fig.~\ref{fig1}, due to the following relations\footnote{$m_u, m_d, m_{\pi}$ are masses of u-quark, d-quark and pion, respectively; $f_{\pi}$ is pion decay constant.}:
\begin{equation}\label{4}
g_{a\gamma \gamma}\; = \; g^0_{a\gamma \gamma} - \frac{e^2}{12\pi^2 f_a}\cdot \frac{4m_d + m_u}{m_u + m_d} \, , \quad m_a = \frac{m_{\pi} f_{\pi} \sqrt{m_u m_d}}{\left( m_u + m_d \right) f_a}\, .
\end{equation}

Let us now elaborate why the construction presented in the previous paragraph contains an implicit far-reaching assumption. While allowing the new heavy fermion $\psi$ to be charged under any possible gauge symmetry at hand is indeed the most generic option, it is not consistently implemented in the KSVZ-like models. The reason is that these models consider only electric representations of the gauge groups. Meanwhile, as it was shown back in 1931 by Dirac~\cite{Dirac:1931kp}, gauge interactions in the quantum theory need not be limited to the electric ones: gauge charges can be electric as well as magnetic.  Although we do not see magnetic charges at low energies, their existence is actually indirectly evidenced by the quantization of the electric charge observed in nature. Indeed, as Dirac found, quantum theory requires that the electric gauge charge $e$ is related to the magnetic one $g$ as follows: $eg = 2\pi n \, , n \in \mathbb{Z}$, so that the charges are quantized. Moreover, as it was advocated in ref.~\cite{Polchinski:2003bq}, there is a mounting evidence supporting the conjecture that charge quantization is not only a necessary but also a sufficient condition for the magnetic monopoles to exist. Quantum field theory coupled to gravity suggests that any possible electric or magnetic charge should have a physical realization. The construction of KSVZ-like models is thus too restrictive, so that the gauge charges assigned to the heavy fermion $\psi$ are not generic and the resulting predictions for axion couplings are biased.

\section{Axion-photon coupling from the axial anomaly of magnetic currents}

Let us now relax the above-mentioned assumption on the representations of the heavy fermion $\psi$ under the gauge groups of the Standard model and consider a truly generic setting. In this setting, $\psi$ is a dyon, i.e. a particle carrying both electric and magnetic charges. We limit our investigation to axion phenomenology at low energies, so that the relevant gauge interactions are electromagnetic and color ones. For simplicity, we also do not consider here the case where $\psi$ has a color magnetic charge, which is studied in detail in our publication~\cite{Sokolov:2021ydn}. Since this means we do not modify the strong sector of the model, phenomenology of the strong interactions of axions is fully analogous to the one in the KSVZ model: in particular, the strong CP problem is solved and the relation between axion mass $m_a$ and decay constant $f_a$ is standard. What is modified are axion-photon interactions. Let us proceed to derive the latter from the UV theory with the Lagrangian given by eq.~\eqref{1}. As it was shown by Zwanziger~\cite{PhysRevD.3.880}, a local quantum field theory with both electric and magnetic charges necessarily involves two vector-potentials, $A_{\mu}$ and $B_{\mu}$, each having the standard coupling to the corresponding current, electric or magnetic, respectively. The covariant derivative entering eq.~\eqref{1} is thus $D_{\mu} = \partial_{\mu} - e_0\, q_e A_{\mu} - g_0\, q_m B_{\mu} $, where $e_0 = e/3$ and $g_0$ are elementary electric and magnetic charges, respectively. Due to the Dirac quantization condition, $g_0 = 2\pi / e_0$. 

\begin{figure}[t]
\vspace*{-0.4cm}
	\begin{floatrow}
		\ffigbox{\caption{Axion-photon coupling as a function of axion mass and decay constant for KSVZ-like and DFSZ axion models together with the existing constraints on the corresponding parameter space from experiments as well as from astrophysical data. Astrophysical hints are also shown.}\label{fig1}}
		{\includegraphics[height=6cm]{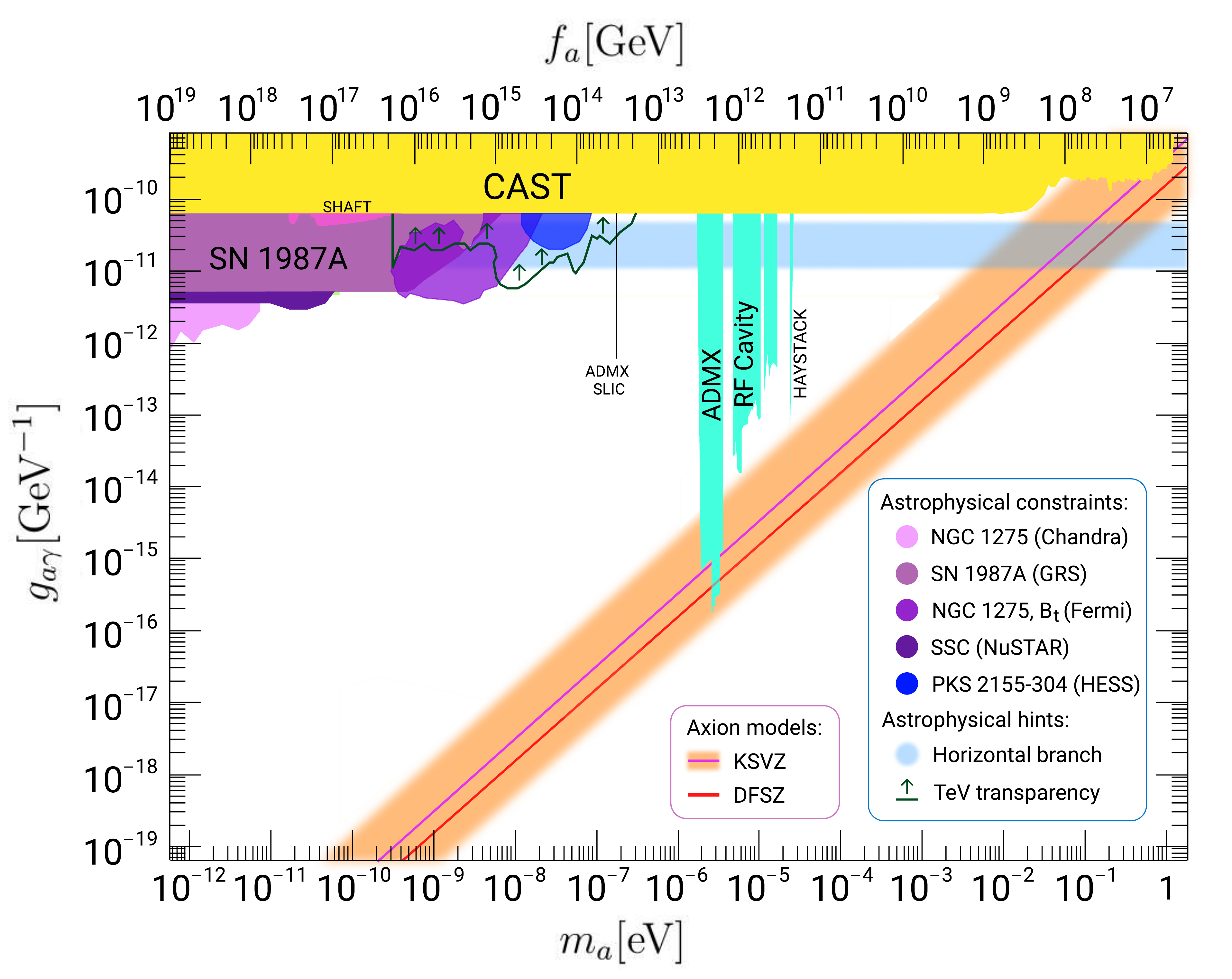}}
		\ffigbox{\caption{Same as fig.~\ref{fig1}, but extended with the projected constraints from future experiments (dashed lines) and  with the blue line corresponding to the model which accounts for the magnetic anomaly coefficients (for a particular case of single new fermion $\psi$ with $q_m = 1\, ,\; d\!\left( C_{\psi} \right) = 3\,$).}\label{fig2}}
		{\includegraphics[height=6cm]{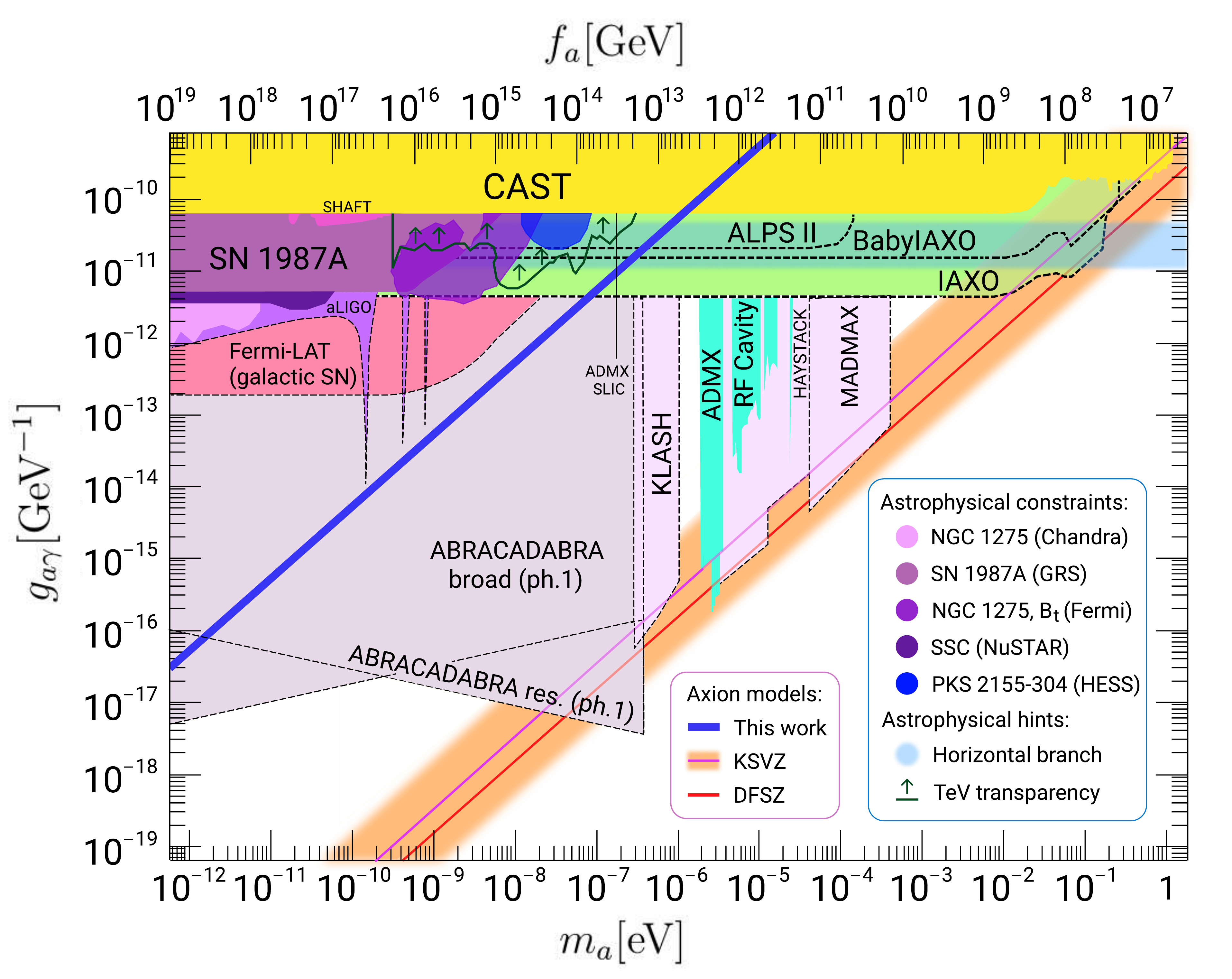}}
	\end{floatrow}
\end{figure}

Below the PQ symmetry breaking scale, one can expand $\Phi = (v_a + \rho ) \exp{\left( i a/v_a \right) } /\sqrt{2}$, where $\rho$ is a heavy radial mode and $a$ is a pseudo Nambu-Goldstone boson (axion). The terms in the resulting Lagrangian which are relevant for the low energy phenomenology are:
\begin{equation}
\mathcal{L} \; \supset \; i\bar{\psi} \gamma^{\mu} D_{\mu} \psi + \frac{y v_a}{\sqrt{2}} \left\lbrace \exp{ \left( \frac{i a}{v_a} \right) }\, \bar{\psi}_L \psi_R + \text{h.c.} \right\rbrace .
\end{equation}
We perform then an axial rotation of the fermion $\psi \rightarrow \exp{(i a\gamma_5 / 2v_a)}\cdot \psi$, after which there arise an anomalous term $\mathcal{L}_{ \text{F}}$ from the transformation of the measure of the path integral and a derivative coupling of $a$ to the axial current of $\psi$:
\begin{equation}\label{6}
\mathcal{L} \; \supset \; i\bar{\psi} \gamma^{\mu} D_{\mu} \psi + \frac{y v_a}{\sqrt{2}}\, \bar{\psi} \psi - \frac{\partial^{\mu} a}{2v_a}\, \bar{\psi} \gamma_{\mu} \gamma_5 \psi - \mathcal{L}_{ \text{F}} \, .
\end{equation}
The axial anomaly in a theory with dyons was found in ref.~\cite{Csaki:2010rv}, from where one infers the expression for the anomalous term\footnote{Although it may naively seem that the axion shift symmetry $a \rightarrow a + 2\pi v_a$ forbids some of the terms in eq.~\eqref{7}, it actually does not, which we elaborate in a forthcoming paper.}:
\begin{equation}\label{7}
\mathcal{L}_{ \text{F}}\; = \; \frac{a\cdot d \! \left( {{C}}_{{\psi}} \right) }{16\pi^2 v_a} \cdot \left\lbrace \left( e_0^2  q_e^2 - g_0^2\, q_m^2 \right) F_{\mu \nu} \tilde{F}^{ \mu \nu} -4\pi q_m q_e F_{\mu \nu} {F}^{ \mu \nu} \right\rbrace  .
\end{equation}
The dimension $d({{C}}_{{\psi}})$ of the color representation of $\psi$  in the numerator on the right-hand side comes from summing over the color indices. We corrected a mistake which crawled into the coefficient of the last term on the right-hand side in the original expression for the anomalous current in ref.~\cite{Csaki:2010rv}. This term is CP-violating and arises only if there is CP-violation in the UV theory. A generic theory with dyons does violate CP, unless for any dyon with charges ($q_e$, $q_m$) there is another dyon with charges ($q_e$, $-q_m$). Anyway, be there CP-violation or not, it is easy to see that the second term on the right-hand side of eq.~\eqref{7} dominates over all other terms: if we assume $q_e, q_m \sim 1$, its coefficient is by a factor of $4\pi^2/e_0^4$ larger than the coefficient in front of the first term and by a factor of $\pi/e_0^2\,$ -- than the coefficient in front of the third term. At the leading order, as long as $m = yv_a/\sqrt{2}$ is very large, the derivative axion coupling from eq.~\eqref{6} does not contribute to the low energy Lagrangian of axion-photon interactions, for the same reason as in the KSVZ-like models\footnote{The reason is the behaviour of the Ward identities for the correlation functions which involve the axial current of $\psi$ in the limit $m\rightarrow \infty$.}. The leading effect in the interactions between axion and photons is thus given by the following Lagrangian:
\begin{equation}
\mathcal{L}_{a\gamma} = \frac{1}{4}\, \tilde{g}_{a\gamma \gamma} \,a F_{\mu \nu} \tilde{F}^{\mu \nu} , \quad \tilde{g}_{a\gamma \gamma} = \frac{q_m^2\, d\! \left( C_{\psi} \right) g_0^2}{4\pi^2 v_a}\, ,
\end{equation}
where we took into account that the contribution to the axion-photon coupling $\tilde{g}_{a\gamma \gamma}$ from the strong sector, see eq.~\eqref{4}, is now absolutely negligible. The axion-photon coupling can be rewritten in terms of the magnetic anomaly coefficients $\tilde{E}_{\psi}$:
\begin{equation}
\tilde{g}_{a\gamma \gamma} = \frac{\tilde{E}}{N} \cdot \frac{g_0^2}{8\pi^2 f_a} \, , \quad \tilde{E} = \sum_{\psi} \tilde{E}_{\psi} = \sum_{\psi} q_m^{ 2} (\psi)\cdot d\! \left( C_{\psi} \right) \, ,
\end{equation}
where $q_m (\psi) \in \mathbb{Z}$ are magnetic charges of new heavy fermions $\psi$. For example, in case of a single new heavy quark $\psi$ with minimal magnetic charge $q_m = 1$ the axion-photon coupling is $\tilde{g}_{a\gamma \gamma} = 3/(e_0^2 f_a) = 27/(e^2f_a)$, which is plotted against axion mass $m_a$ and decay constant $f_a$ in fig.~\ref{fig2}. This result is consistent with the one we obtained using the Schwinger proper time method in ref.~\cite{Sokolov:2021ydn}. In the latter article, we also discuss further phenomenological implications, including axion-nucleon couplings and axion dark matter for pre-inflationary PQ symmetry breaking.

\section*{Acknowlegments}
	A.R. acknowledges support and A.S. is funded by the Deutsche Forschungsgemeinschaft (DFG, German Research Foundation) under Germany's Excellence Strategy -- EXC 2121 \textit{Quantum Universe} -- 390833306.

\bibliographystyle{JHEP}
{\small \bibliography{PoS}}

\end{document}